\documentclass[12pt, prd, showpacs]{revtex4}
\usepackage{amssymb}
\usepackage{amsmath}

\begin{document}

\title{Unified approach to redshift in cosmological /black hole spacetimes
and synchronous frame}
\author{A.  V. Toporensky}
\affiliation{Sternberg Astronomical Institute, Lomonosov Moscow State University,
Universitetsky Prospect, 13, Moscow 119991, Russia }
\affiliation{Kazan Federal University, Kremlevskaya 18, Kazan 420008, Russia}
\email{atopor@rambler.ru}
\author{O. B. Zaslavskii}
\affiliation{Department of Physics and Technology, Kharkov V.N. Karazin National
University, 4 Svoboda Square, Kharkov 61022, Ukraine}
\affiliation{Kazan Federal University, Kremlevskaya 18, Kazan 420008 Russia}
\email{zaslav@ukr.net}
\author{S. B. Popov}
\affiliation{Sternberg Astronomical Institute, Lomonosov Moscow State University,
Universitetsky Prospect, 13, Moscow 119991, Russia }
\email{polar@sai.msu.ru}

\begin{abstract}
Usually, interpretation of redshift in static spacetimes (for example, near
black holes) is opposed to that in cosmology. In this methodological note we
show that both explanations are unified in a natural picture. This is
achieved if, considering the static spacetime, one (i) makes a transition to a
synchronous frame, and (ii) returns to the original frame by means of local
Lorentz boost. To reach our goal, we consider a rather general class of
spherically symmetric spacetimes. In doing so, we construct frames that
generalize the well-known Lemaitre and Painlev\'{e}--Gullstand ones and
elucidate the relation between them. This helps us to understand, in an
unifying approach, how gravitation reveals itself in different branches of
general relativity. This framework can be useful for general relativity university
courses.
\end{abstract}

\maketitle

\section{Introduction}

Though the calculation of redshifts in General Relativity (GR) has no principal
difficulties, their interpretation continues to be a source of debates. It
is well known that gravitational field can induce a blue/red shift in cases which do not have 
counterparts  in Special Relativity (SR). For example, a photon emitted near a
black hole horizon reaches a distant observer severely redshifted, even in
cases where emitter and observer do not move with respect to each other. The gravitational redshift of this type can be
obtained as follows. In the static field, one can introduce such a time
variable $t$ whereby different static observers measure the same time
intervals $dt$ between a pair of corresponding events (say, emission and
absorption of photons). Then, the intervals of the proper time  differ by
the ratio of the factors $\sqrt{g_{tt}}$, where $g_{tt}$ is the
corresponding component of the metric tensor. One can also describe this
redshift in a more formal fashion. In the static gravitation field, there
exists a timelike Killing vector $\xi ^{\mu }$ (see, for example, \cite{wein}%
). Let a photon have the wave vector $k^{\mu }$. Then, the quantity $\omega
_{0}=-k_{\mu }\xi ^{\mu }$ is conserved along the path of a photon due to
the geodesic equation for $k^{\mu }$ and Killing equations. This has the
meaning of a frequency measured at infinity. Let us imagine a set of static
observers with the four-velocities $u^{\mu }$; they measure the local
frequencies $\omega =-k_{\mu }u^{\mu }$. For a static observer, $u^{\mu }={%
\xi ^{\mu }}/{N}$, where $N=\sqrt{-\xi _{\mu }\xi ^{\mu }}$ is a lapse
function. Then, we obtain $\omega N=\omega _{0}$ which gives a redshift
in a static field. These two methods imply staticity of the field and are
completely equivalent.

However, the cosmological redshift does not fall into this scheme. In an
expanding universe, one is tempted to turn to explanation in the
spirit of the Doppler effect. However, it cannot be simply reduced to this
effect (at least globally, see discussion below) since, in particular,
coordinate recession velocities can exceed the speed of light. Therefore, in GR, one cannot use straightforward
generalization of the Doppler effect in SR which is based on the treatment of physical velocities of the observer and emitter.
Meanwhile, we cannot calculate the redshift as it is done in the black hole
spacetime because in expanding or contracting Universe there is no timelike
Killing vector. Gravitational time delay is also absent in the
Friedmann-Robertson-Walker (FRW) coordinate system because this system is
synchronous, which means that all particles in the Hubble flow share the
same time that coincides with the proper time. For this reason, the
cosmological redshift is sometimes treated as a third type of redshift
which is distinct from both the Doppler and gravitational ones (see, e.g.
Sec. 15 of \cite{har}).

Such a classification of redshifts, though being rather useful from a practical point
of view (since it prevents the use of incorrect formulae for inappropriate
types of redshifts), has its own conceptual problems. The reason for this is, in
this picture, cosmology appears to be separated from other GR issues (which
seems rather strange), and more importantantly, such a division cannot be
done technically in many situations which are slightly more sophisticated
than the FRW metrics. For example, should redshifts in the Schwarzschild-de
Sitter solution  be interpreted in cosmological terms, or it is better
to apply terminology from black hole physics?

The goal of the present methodological paper is to show that apparent
``cosmological'' features of cosmological redshift are not limited only by
situations related to cosmology, but, rather, follow from properties of the
synchronous coordinate system (for example, the FRW metric), and therefore  can be
present in any other applications of GR (including black holes) when an
appropriate coordinate system is chosen. We think that any graduate student
who has chosen General Relativity for a deeper study, as well as young
researcher starting to work in this field, should be able to interpet
physical phenomena in different frames and understand the relationship
between them (our recent experience of supervising graduate students
confirms this), and we hope that this paper will help to reach this goal. This
ability is especially needed while describing effects connecting with the
black hole horizon crossing, because coordinate singularity of commonly used
stationary coordinate system has sometimes induced incorrect statements. One of
the most famous example is the concept of ``ghosts'' emitted at a black hole
horizon, seen from the zero distance while an observer crosses the horizon 
\cite{ghost1,ghost2}. This concept has already been criticized in \cite{Kas}. We will
comment on this point later on.

There are  other unifying approaches to redshifts, one of which is based
on parallel transport of the observer's velocity along the lightlike
geodesic \cite{narl}, \cite{synge}, \cite{dop}. In contrast, we do not
introduce additional constructions and operate with equations of motion
directly. Throughout the paper, we use many concepts from GR and relativistic
cosmology. We suggest the use of standard textbooks \cite{wein, mtw}
as references books in GR.

\section{Cosmological redshift}

Derivation of the cosmological redshift is a source of controversy ---
different authors follow different routes to obtain this quantity. We
briefly mention the most popular of them, and then present another one,
which will be used in the rest of the paper.

We start with the usual expression for the interval~\footnote{%
Only radial motions are discussed.} (see eq. 14.2.1 in \cite{wein}):

\begin{equation}
ds^{2}=-c^{2}dt^{2}+a^{2}(t)d\chi ^{2}.  \label{m}
\end{equation}%
Here $c$ is the light speed. Hereafter, we assume radial motion only and suppress angular cooridnates.

The simplest way to obtain the redshift in the FRW metric is to use the
conformal time, $\eta $. Defining $\eta (t)=\int dt/a$ we reduce the metric
to the form%
\begin{equation}
ds^{2}=a^{2}(d\chi ^{2}-d\eta ^{2}).  \label{m2}
\end{equation}%
Assuming $\chi _{obs}=0$ we obtain $\chi _{em}=\eta _{obs}-\eta _{em}$.
Here, subscripts ``obs'' and ``em'' refer to an observer and emitter,
respectively. If a signal was emitted during a period $\Delta \eta $, then
it is observed during the same interval of conformal time. However, physical
time intervals at two moments are different: $\Delta t_{em}=a(t_{em})\Delta
\eta $ and $\Delta t_{obs}=a(t_{obs})\Delta \eta $. Thus, $\Delta
t_{em}/\Delta t_{obs}=a(t_{em})/a(t_{obs})$, where we neglected the change
of the scale factor during the processes of emission and observation.

Then, one can use symmetry of a metric to derive the redshift using
conservation laws (see pp. 457-458 of \cite{mtw}). It is possible to show
that in the FRW metric the momentum changes as $a(t)^{-1}$  (see, e.g. eq. 114.5 in Ref. \cite{ll}).
 Then, as for
photons momentum $\sim \lambda ^{-1}$, we obtain that $\lambda \sim a(t)$%
. Thus, we derive the usual expression for the redshift.

Both of these methods exploit properties of a metric (in particular, FRW),
and so they are not general derivations.

The third way we want to mention is popular, however, it is not fundamental,
as \textit{a priori} it is not obvious that it might provide the correct
answer. This method uses integration of the local Doppler effect along the
trajectory of a photon, or even just the statement that locally
cosmological redshift can be given by the Doppler formula. Such a simplified
approach, most often with appropriate comments that this is not an accurate
 but just illustrative  way of derivation, can be found in some
textbooks (for example, Ch. 5.2 of \cite{ldl}).

The most general method one can often find in textbooks uses null geodesics
(see pp. 777 - 778 of \cite{mtw}).

For photons

\begin{equation}
ds^{2}=0.  \label{s0}
\end{equation}

Thus, we obtain from (\ref{m}) the expression for the coordinate velocity:

\begin{equation*}
d\chi/dt=c/a.
\end{equation*}
Let us then consider two pulses of an electromagnetic wave. They are
emitted in two moments of time $t_{em1}$, $t_{em2}$ and observed,
correspondingly, at $t_{obs1}$ and $t_{obs2}$. Both are emitted by a galaxy
at $\chi _{em}$. The observer is situated at $\chi _{obs}$.

If an emitter and an observer follow the geodesic lines, their values of $\chi $
remain fixed since (\ref{m}) represents a synchronous system. World lines of
the two photons are:

\begin{equation}
\chi _{em}-\chi _{obs}=c\int_{t_{em1}}^{t_{obs1}}a^{-1}dt,  \label{1}
\end{equation}%
\begin{equation}
\chi _{em}-\chi _{obs}=c\int_{t_{em2}}^{t_{obs2}}a^{-1}dt.  \label{2}
\end{equation}

Then:

\begin{equation}
\int_{t_{em1}}^{t_{obs1}}a^{-1}dt-\int_{t_{em2}}^{t_{obs2}}a^{-1}dt=0.
\label{12}
\end{equation}

Eq. (\ref{12}) can be rewritten in the equivalent form%
\begin{equation}
\int_{t_{obs1}}^{t_{obs2}}a^{-1}dt-\int_{t_{em1}}^{t_{em2}}a^{-1}dt=0\text{.}
\end{equation}

Neglecting changes in $a$ during the interval of emission of the two pulses,
and during the interval of their observation (i.e. putting $%
a(t_{em1})=a(t_{em2}) $ and $a(t_{obs1})=a(t_{obs2})$), we obtain

\begin{equation*}
\frac{t_{obs2}-t_{obs1}}{a(t_{obs})}-\frac{t_{em2}-t_{em1}}{a(t_{em})}=0.
\end{equation*}

At emission $\lambda_{em}=c(t_{em2}-t_{em1})$. At observation $%
\lambda_{obs}=c(t_{obs2}-t_{obs1})$. Therefore,

\begin{equation*}
\lambda_{em}/a(t_{em})=\lambda_{obs}/a(t_{obs}).
\end{equation*}
This corresponds to equations (14.3.5), (14.3.6) in \cite{wein}). Finally, we
come to the usual expression for the cosmological redshift.

In our paper, we use a variation of this method. In our opinion, it is
interesting from the methodological point of view and also
provides a clear visualization of the process. For our goals here, this
version of derivation is needed because, unlike the preceding method, it can be
easily generalized to inhomogeneous metrics, which will be used below. Our method
essentially relies on using different definitions of velocity; this contains
some subtleties discussed below.

\section{Definitions of velocity and ``light faster than light''}

We note that the fact that the coordinate velocity of light, $d\chi /dt$,
can be not equal to $c$ in some coordinate systems is obvious, because
coordinates have no direct physical meaning. What is less obvious is that
the velocity of light, defined as derivative of the proper distance from an
observer to a photon over proper time (which in the FRW metric is equal to
the coordinate time), can be different from $c$ as well.

To work with proper distance, we need to integrate $\sqrt{g_{\chi \chi }}$
over $d\chi $ at the same moment of proper time. This means that we have to
define hypersurfaces $t=const$. To do this, it is necessary to fix a
coordinate system; then the answer depends on the coordinate system, i.e.
is not invariant.

In the FRW metric (\ref{m}) the proper distance between $\chi =0$ and some $%
\chi $ at a given moment of time $t$ is equal to $r=\chi a$. If $\chi ~$is
allowed to vary with time, for the velocity $dr/dt$ we can write a chain of
equations:

\begin{equation}
\frac{dr}{dt}=\frac{d(\chi a)}{dt}=\frac{d\chi }{dt}a+\frac{da}{dt}\chi =c+%
\frac{da}{dt}\chi =c+\frac{\dot{a}}{a}a\chi =c+v_{fl},  \label{rt}
\end{equation}%
where $v_{fl}=Hr$ is the Hubble flow velocity and $H={\dot{a}}/{a}$ is the
Hubble constant; the dot denotes differentiation with respect to $t$. This clearly demonstrates so-called \textquotedblleft
superluminal motion\textquotedblright\ in cosmology. Note that eq. (\ref{rt}%
) can be considered as a particular case of the general fact that, in
cosmology, peculiar velocity is summed with the Hubble velocity following the
Galilean law. It is worth stressing that a local velocity of light defined
as $a{d\chi }/{dt}$ (the first term in eq. \ref{rt}) is equal to $c$
according to (\ref{s0}), as it should be.

To make the difference between two definitions of velocity more pronounced,
let us denote $P(\chi _{2},t\mid \chi _{1},t)$ as the proper distance between
points 1 and 2 at the moment $t$. Then,%
\begin{equation}
\frac{dr}{dt}=\frac{dP(\chi ,t\mid 0,t)}{dt}=v_{loc}+v_{fl},  \label{rp}
\end{equation}%
where the local velocity%
\begin{equation}
v_{loc}=\lim_{\Delta t\rightarrow 0}\frac{P(\chi +\Delta \chi ,t\mid \chi ,t)%
}{\Delta t}=\frac{\partial P(\chi ,t\mid 0,t)}{\partial \chi }\frac{d\chi }{%
dt}\text{,}  \label{vloc}
\end{equation}%
and the flow velocity%
\begin{equation}
v_{fl}=\frac{\partial P(\chi ,t\mid 0,t)}{\partial t}\text{.}  \label{vfl}
\end{equation}

It is the quantity $v_{loc}$ that can be found by pure local measurements.
For light, $v_{loc}=c$. The quantity ${dr}/{dt}$ is more involved and cannot
be detected by direct local measurements. Not surprisingly, it can exceed $c$
without any contradiction with GR and SR.

Now let us derive the equation for the cosmological redshift. Here we will
use the fact that two light pulses, being at slightly different distances
from an observer, have different velocities with respect to this observer.
Therefore, let us consider two neighboring maxima of a light wave directed toward
an observer, separated by a spatial distance $\Delta r=\lambda $. This
radial difference results in the velocity difference with respect to the
observer $\Delta v=H\Delta r$. This means that the velocity of the maximum
which is closer to the observer is less than the velocity of the maximum
which is behind it. Due to this difference, the distance between these
maxima changes and we can construct a differential equation 
\begin{equation}
d\Delta r/dt=H\Delta r.  
\label{H}
\end{equation}%

After obvious calculations using the definition of the Hubble parameter $H=%
\dot{a}/a$ we obtain that the ratio of wavelengths at $t_{em}$ and the
moment of observation, $t_{obs}$, is equal to the inverse ratio of scale
factors at these moments: $\lambda _{em}/\lambda _{obs}=a_{em}/a_{obs}$.
This is the standard result for the cosmological redshift.

Below, we want to show how features typical for cosmology can be considered
in a rather generic black hole background if a synchronous system is used.
In doing so, we generalize the well-known Lemaitre metric connected to an
observer free-falling into a black hole. Then, it is clear that the concept
of an expanding or contracting space (just discussed in the cosmological
context) by itself is not necessary to understand the physics of
cosmological evolution.

First of all, the definition of the Hubble flow and peculiar velocity has a
straightforward generalization. Indeed, in a synchronous system coordinate
lines $x^{i}=const$ are geodesics (see, e.g. Sec. 11.97 of \cite{ll}).
Observers with constant spatial coordinates share the same time coordinate
which coincides with their proper time. One can define a proper distance
between two points at some particular time $t$ and also velocity of one
point with respect to another by differentiating the proper distance with
respect to $t$. For simplicity, we will consider spherically symmetric
metrics.

Let $\chi $ be the radial coordinate,   $\chi _{2}$ be its value of the observed particle which
changes in time, while the coordinate $\chi _{1}$ of the observer remains fixed.
Then we get the following, using the well-known formula for the integral with varying
limits:

\begin{equation}
v=\frac{d}{dt}\int_{\chi _{1}}^{\chi _{2}}\sqrt{g_{\chi \chi }}d\chi
=v_{fl}+v_{loc}\text{,}  \label{sum}
\end{equation}

The velocity appears to be a sum of two components. The first one, 
\begin{equation}
v_{fl}=\int_{\chi _{1}}^{\chi _{2}}\frac{d}{dt}\sqrt{g_{\chi \chi }}d\chi
\label{bhf}
\end{equation}%
is fully due to the non-static character of the metric (and can be
considered as an analogue of velocity of the Hubble flow). The second
one, 
\begin{equation}
v_{loc}=\sqrt{g_{\chi \chi }(\chi _{2})}d\chi _{2}/dt  \label{bhl}
\end{equation}%
is non-zero only when the radial coordinate of a particle changes in time,
so we can treat it as an analogue of peculiar velocity in cosmology. Eqs. (%
\ref{sum}) -- (\ref{bhl}) represent the counterpart of cosmological eqs. (\ref{rt}) -- (%
\ref{vfl}). Of course, for light locally $v_{loc}$ is always
equal to $c$.

\section{Static spacetime and relationship between frames: Lemaitre vs.
Painlev\'{e}--Gullstand metrics}

In what follows, we consider a static metric in a \textquotedblleft
cosmological manner\textquotedblright . This includes the black hole
spacetime. We show similarity between phenomena in black hole physics and
cosmology, and demonstrate that, looking different at the first glance,
these processes have the same interpretation in a synchronous metric. We
discuss a black hole metric of a rather general form. Without the loss of
generality, we take it to be 

\begin{equation}
ds^{2}=-f(r)dt^{2}+\frac{dr^{2}}{f(r)}+r^{2}(d\theta ^{2}+\sin ^{2}\theta
d\phi ^{2})\text{.}  \label{mf}
\end{equation}%
Hereafter, we put the fundamental constant $c=1$.
This equation includes the Schwarzschild metric ($f=1-{r_{g}}/{r})$, as well as the
Reissner-Nordstr\"{o}m one, etc. Extension to the case $g_{00}g_{11}\neq -1$
is straightforward. The event horizon lies at $r=r_{g}$, so $f(r_{g})=0$.
Near the horizon, the metric becomes degenerate; this can be remedied by
introducing a new set of coordinates.

Making the coordinate transformations \cite{along}

\begin{equation}
\rho =t+\int \frac{dr}{f\sqrt{1-f}}\text{,}  \label{rof}
\end{equation}%
\begin{equation}
\tau =t+\int \frac{dr}{f}\sqrt{1-f}  \label{tauf}
\end{equation}%
one can reduce the metric to the form

\begin{equation}
ds^{2}=-d\tau ^{2}+(1-f)d\rho ^{2}+r^{2}(\rho ,\tau )(d\theta ^{2}+\sin
^{2}\theta d\phi ^{2})\text{.}  \label{metL}
\end{equation}%
If $f=1-{r_{g}}/{r}$, we return to the well-known Lemaitre metric \cite{ll}.
Therefore, we will call (\ref{metL}) the generalized Lemaitre (GL) frame.

Since the GL metric is synchronous, all free falling radial observers have
the constant comoving radial coordinate $\rho $ and share the same proper
time. It appears that the proper distance is related with $r$ in a very
simple way. Namely, fixing $\tau $, one finds from (\ref{rof}) -- (\ref{metL}%
) after simple calculations that the proper distance between points 1 and
2 equals%
\begin{equation}
l=\int_{\rho _{1}}^{\rho _{2}}d\rho \sqrt{1-f}=r_{2}-r_{1}\text{.}
\label{dis}
\end{equation}%
In what follows, we will use $r$ along with the radial coordinate $\rho $.
We can see here some analogue of the cosmological situation with the FRW
metric where the proper distance between two points is given by the formula $%
d=a(\chi _{2}-\chi _{1})$, which is valid for any observer moving in the
Hubble flow.

It is worth noting that for massive particles composing the GL system $E=m$, 
$\rho =const$ and thus $v_{fl}=v$, $v_{loc}=0$. For a photon $\left\vert
v_{loc}\right\vert =1$.

One can also make one more step and write down the metric in a ``mixed''
form using a new time variable $\tau $ and old radial variable $r$. Then,
eliminating $dt$ we have from (\ref{rof}), (\ref{tauf}):%
\begin{equation}
ds^{2}=-d\tau ^{2}f+dr^{2}+2drd\tau \sqrt{1-f}+r^{2}(d\theta ^{2}+\sin
^{2}\theta d\phi ^{2})\text{.}  \label{pg}
\end{equation}

In the Schwarzschild case, the metric (\ref{pg}) is nothing else but the
so-called Painlev\'{e}--Gullstand metric (see, e.g. \cite{regular} and
references therein). For a generic $f$, it is natural to call it the
generalized Painlev\'{e}--Gullstand metric (GPG). It is worth noting that it is
metric (\ref{pg}) which was used in \cite{river} to describe
\textquotedblleft a river model of black hole\textquotedblright\ as an
interesting counterpart of the cosmological concept of \textquotedblleft
expanding space\textquotedblright . Strange as it may seem, the relationship
between the Painlev\'{e}--Gullstand and Lemaitre metrics, to the best of our
knowledge, was not noticed before.

In these coordinates, some properties of the metric look rather simple. In
particular, the hypersurfaces $\tau =const$ are flat. Eq. (\ref{dis})
becomes a direct consequence of (\ref{pg}). For a massive particle falling
along geodesic from the rest at infinity, one has ${dr}/{d\tau }=-\sqrt{1-f}$%
, whence we  have for the velocity $v=\left\vert {dr}/{d\tau }\right\vert $:

\begin{equation}
v=\sqrt{1-f}\text{.}  \label{vf}
\end{equation}%
If points 1 and 2 move along geodesics, their relative velocity is:%
\begin{equation}
v_{12}=v_{2}-v_{1}\text{.}  \label{v12}
\end{equation}

For light propagation we have:%
\begin{equation}
\frac{dr}{d\tau }=\pm 1-v\text{,}  \label{l}
\end{equation}%
where $v$ coincides with (\ref{vf}). Here, sign ``+''
corresponds to propagation outward and ``-'' corresponds to propagation
inward.

\section{Behavior of wavelength}

Now, by the same method as in the FRW cosmology, we derive the redshift of a
photon emitted by a free falling particle at $r_{em}$ and absorbed by
another free falling particle at $r_{obs}$ . Remember that the redshift in
the FRW frame appears because the proper distance between two light pulses
changes with time since the speed of light with respect to a distant
observer depends on the distance between the pulse and the observer. The
same situation appears in the Lemaitre metric, as we will see soon. Our goal
is to start with two maxima of light wave separated by $\Delta r=\lambda
_{1} $ and trace the evolution of this separation during light propagation.
An observer will see the maxima separated by $\lambda _{2}$ which gives us
the redshift via $1+z=\lambda _{2}/\lambda _{1}$. As usual for synchronous
systems, the coordinate velocity of light is $v_{fl}+c$. Hereafter, we put c=1 for simplicity. 
For two points with
radial difference $\Delta r$ at $r_{em}$ (here, the dependence on radial
coordinate at emission is important, because the metric is not homogeneous)
corresponding velocities of light differ by $\Delta v_{fl}=v_{fl}^{\prime
}(r)\Delta r.$ In this Section, we consider free-falling particles co-moving
with respect to the GL system, so peculiar velocities vanish and $v_{fl}=v$.

It is worth noting that while we need to subtract the velocity of the
observer in order to get relative velocity (\ref{v12}), this does not affect
the \textit{difference} between velocities of nearby points $\Delta v$, so
we can consider the position $r_{0}$ of the observer as arbitrary. For
simplicity, we consider free-fall velocities with respect to an infinitely
distant observer.

Thus, if at emission the distance between two wave maxima was $\left\vert
\Delta r\right\vert \equiv \lambda $, it later increases according to%
\begin{equation}
d\Delta r=\Delta vd\tau =v^{\prime }(r)\Delta rd\tau,  \label{lat}
\end{equation}%
if $v$ depends upon $r$ only (this is true for
all static spherically symmetric spacetimes).
This equation is an analogue of the cosmological equation (\ref{H}). It is
useful to integrate over the radial coordinate $r$ (instead of the Lemaitre
coordinates $\rho $ and $\tau $). Let us consider an inwardly moving photon
so that, according to (\ref{l}), $d\tau =-{dr}/{(1+v)}$. Then we get:

\begin{equation}
\frac{d\lambda }{dr}(1+v)=\frac{dv}{dr}\lambda ,  \label{dla}
\end{equation}%
and finally

\begin{equation}
\frac{d\lambda}{\lambda}=\frac{dv}{1+v}.
\end{equation}%
Note, that in the FRW cosmology $v$ is not a function of time only in
the de Sitter universe where $dv/dr=H=const$.

Since for the photon moving inward, the sum $v+1$ never crosses zero, the
differential equation is regular at the event horizon, so the solution
should give us the redshift of such a photon for both possible locations of
an emitter and an observer (inside or outside the event horizon).

Then, we have the ratio of the initial and final wavelengths for an inward
motion of a photon:

\begin{equation}
\frac{\lambda _{obs}}{\lambda _{em}}\equiv 1+z=\frac{1+v_{obs}}{1+v_{em}}.
\label{z}
\end{equation}

For the outward motion of a photon, in a similar way one would obtain%
\begin{equation}
1+z=\frac{1-v_{obs}}{1-v_{em}}\text{.}  \label{out}
\end{equation}

These results are in agreement with those of \cite{ras2013} and \cite{Kas}
(in the latter case they were derived with the help of the Kruskal
coordinates). As noted above, in FRW cosmology, these formulas are valid only
for the de Sitter solution (see more comments on this in the following).
 For other regimes, the dependence between the
redshift and the velocity at emission (in FRW metrics, it is reasonable to
set the velocity of observer to zero) is more complicated; see, for example, 
\cite{our1}.

Going back to static spacetimes, we can use (\ref{vf}) to get the general
expression for the redshift through radial coordinate of emission and
observation: 
\begin{equation}
1+z=\frac{1\pm \sqrt{1-f(r_{obs})}}{1\pm \sqrt{1-f(r_{em})}}  \label{zr}
\end{equation}%
where the upper sign is used for ingoing photon, and the lower sign for the
outgoing photon.

The latter case can be applied if observer and emitter are either both
outside the horizon (in this situation evidently $r_{em}(\tau
_{em})<r_{obs}(\tau _{obs})$), or both inside the horizon (and $r_{em}(\tau
_{em})>r_{obs}(\tau _{obs})$). On the contrary, there are no restrictions for
the locations of emitter and observer with respect to the horizon for the ingoing photon case.

If, for example, an observer crosses the horizon, $f_{obs}=0$, then
according to (\ref{vf}), $v_{obs}=1$. Thus, he/she receives a signal from a
remote source at infinity ($f_{em}=1$) with the redshift $z=1$.

If the observer approaches the singularity where $r\rightarrow 0$ and $%
f\rightarrow -\infty $ (like in the Schwarzschild metric where $f=1-{r_{g}}/{%
r}$), $v_{em}\rightarrow \infty $, so the redshift diverges.

It is clear from the very method of derivation that these results are quite
general and are valid far beyond GR. It is necessary, of course, that free
massive particles follow geodesics; this holds in metric theories of
gravity. Also, it is necessary that photons move along null geodesics. This
is achieved if the coupling between electromagnetism and gravity is minimal.
This second condition is fulfilled in most popular generalizations of GR,
though may be violated for some class of theories studied by Horndeski \cite%
{horn}.


It is instructive to illustrate some formulas just derived using the
Schwarzschild metric as an example. This is the most simple and, at the same
time, physically relevant metric describing a black hole. In this case, $f=1-%
\frac{r_{g}}{r}$, where $r_{g}=2M$ is the horizon radius in geometric units
(the fundamental constants $G=c=1$), $M$ being a black hole mass. 



Eq. (\ref{zr}), describing the frequency shift for freely moving particles,
gives us 
\begin{equation}
1+z=\frac{1\pm \sqrt{\frac{r_{g}}{r_{obs}}}}{1\pm \sqrt{\frac{r_{g}}{r_{em}}}%
}\text{.}  \label{1+zs}
\end{equation}

 If a photon
is emitted at infinity and registered when the observer crosses the horizon
(so $r_{em}\rightarrow \infty $ and $r_{obs}=r_{g}$), (\ref{1+zs})  gives
us $z=1$. Thus the frequency is as two times as little as compared to the
original one, in agreement with a general property noted above.

One can ask whether our approach can describe stationary metrics which
are not asymptotically flat? The main example of such a case is the
Schwarzschild-de Sitter metric $f=1-\frac{2M}{r}-\frac{\Lambda r^{2}}{3}$,
where the cosmological constant $\Lambda >0$. In principle, our approach
needs some modifications in this case because we cannot use the condition
that the velocity of free fall is equal to zero in spatial infinity (we
remind a reader that the peculiar velocity is set to zero in the present
paper). We leave a careful treatment of this case to a future work.


However, we can rather easily incorporate into our scheme the pure de Sitter
metric which, in stationary coordinates, has $f=1-\frac{\Lambda r^{2}}{3}$.
In this case, we can demand zero velocity at the coordinate origin (instead
of infinity). This corresponds to the usual definition of zero peculiar
velocity in cosmology. It is worth noting that since the direction of a
\textquotedblleft free fall\textquotedblright\ is changed -- it is now
directed to bigger values of radial coordinate -- the signs in formulae for
inward and outward photons interchange.%

If, in addition to this, we set an observer at the origin of coordinate
system (as usually done in FRW cosmology), we can see that Eq.~(29) for the
redshift is still valid and coincides with the known cosmological result
(see, for example \cite{our1}). As for Eq.(22), it gives us the famous
Hubble law $v=\sqrt{\frac{\Lambda r^{2}}{3}}=Hr$. Obviously, this happens
because the synchronous frame for the de Sitter metric is just the  FRW
frame (1).

\subsection{Special case: propagation along the horizon}

One particular case of an outward propagation of a photon is still beyond
our analysis. Namely, if a photon is emitted exactly at the horizon and
further propagates along it, we cannot integrate over $r$ because it is
constant for such photons (but the Lemaitre coordinate $\rho $ obviously
changes).

Instead, we use directly (\ref{lat}) with $\Delta r$ replaced with $\lambda $
and $r=r_{g}$, whence%
\begin{equation}
d\lambda =\lambda v^{\prime }(r_{g})d\tau \text{,}
\end{equation}%
and $1+z=\exp [v^{\prime }(r_{g}){(\tau _{obs}-\tau _{em})}].$ Using (\ref%
{vf}) and the fact that $f(r_{g})=0$, $f^{\prime }(r_{g})=2\kappa $, where $%
\kappa $ is the surface gravity, we obtain:%
\begin{equation}
1+z=\exp [\kappa {(\tau _{obs}-\tau _{em})}].
\end{equation}

Thus, in this case, the redshift can be expressed through the difference
between emitting and observing time, in agreement with the result obtained in 
\cite{along} by another method. In the particular case of the Schwarzschild
metric, the same problem has been considered earlier using Kruskal
coordinates \cite{Kas}.

Thus, we demonstrated that the Lemaitre coordinates enable us to obtain the
final value of the redshift easily. It is also easy to see that the distance
from the emitter to the observer is finite and well-defined (either at the
time of emission or observation). On the contrary, the radial
coordinate $r$ of emission and observation is the same, which sometimes
leads to a wrong interpretation mentioned in the Introduction. The Lemaitre
coordinates prevent our intuition from making such wrong statements.

\section{Frequencies in different frames}

In this section, we obtain the same results for redshifts in a more formal
way, as it is usually done if problems of interpretations are not concerned.

The geodesic equations for radial light propagation in the static frame are
the following (see Eq.~\ref{mf}):%
\begin{equation}
k^{t}=\frac{\omega _{\infty }}{f}\text{,}  \label{0}
\end{equation}%
\begin{equation*}
k^{r}=-\omega _{\infty }\text{,}
\end{equation*}%
where, for definitness, we consider motion in the inward direction
only. Here, $k^{\mu }$ is the wave vector, $\omega _{\infty }$ has the
meaning of the frequency measured by a remote observer at infinity.

Using (\ref{rof}) and (\ref{tauf}), one can find%
\begin{equation}
k^{\tau }=\frac{\omega _{\infty }}{f}(1-\sqrt{1-f})\text{,}  \label{tauk}
\end{equation}%
\begin{equation}
k^{\rho }=-\frac{\omega _{\infty }}{f\sqrt{1-f}}(1-\sqrt{1-f}).  \label{rok}
\end{equation}

The frequency measured by a local observer with the four-velocity $u^\mu$
is: 
\begin{equation}
\omega =-k_{\mu }u^{\mu }\text{,}  \label{fr}
\end{equation}%
where $u^{\mu }={dx^{\mu }}/{d\tau }$ is the four-velocity and $\tau $ is the
proper time.

For the GPG observers $\rho =const$, so $u^{\tau }=1$, $u^{\rho }=0$. Then,
we have from (\ref{fr}) that%
\begin{equation}
\omega =k^{\tau }=\frac{\omega _{\infty }}{f}(1-\sqrt{1-f})=\frac{\omega
_{\infty }}{1+\sqrt{1-f}}\text{.}  \label{loc}
\end{equation}%
In terms of the velocity (22), it can be rewritten as%
\begin{equation}
\omega =\omega _{\infty }\frac{(1-v)}{f}=\frac{\omega _{\infty }}{1+v}.
\label{omv}
\end{equation}

If light is emitted at $\rho _{em}$ and is absorbed at $\rho _{obs}<\rho
_{em}$,%
\begin{equation}
1+z\equiv \frac{\omega _{em}}{\omega _{obs}}=\frac{1+v_{obs}}{1+v_{em}}
\label{zl}
\end{equation}%
which coincides with (\ref{z}), as it should be. If $\rho _{obs}>\rho _{em}$%
, we again obtain (\ref{out}).

\section{From the GL frame to the static one}

In the above analysis, we have followed the usual cosmological pattern to derive the
redshift. We can also use these results to get the standard redshift in the
Schwarzschild geometry, which is usually obtained using the notation of
gravitational time delay, and is described as a typical example of a
gravitational redshift. This concerns an observer and emitter with a fixed
Schwarzschild radial coordinate $r$. We want to reproduce it using the above
formulae without any references to the gravitational \textquotedblleft time
delay\textquotedblright .\ 

The preceding formulas describe redshift (or blueshift) in the GL frame, when
frequencies or wavelengths are measured by the corresponding free falling
observers who have $\rho =const$. Now we are interested in a redshift in
the static frame. To obtain it, it is sufficient to make a boost from the GL
frame. It follows from (\ref{rof}), (\ref{tauf}) that, for an observer or
emitter to be fixed at a constant $r$, it must have $\rho =\tau +const.$
Then, it moves with the velocity $v={dl}/{d\tau }$ (\ref{vf}) with respect
to the GL frame. Therefore, we need to combine the redshift (\ref{z}) with the
Lorentzian boosts corresponding to the peculiar motion of the emitter and the
observer (i.e. the Doppler effect).

Let us assume that light propagates in the outward direction. Using the
standard formula for the Doppler redshift, we get in the static frame for
the emitter%
\begin{equation}
\left( \omega _{st}\right) _{em}=\left( \omega _{GL}\right) _{em}\sqrt{\frac{%
1-v_{em}}{1+v_{em}}}\text{,}
\end{equation}%
where subscripts \textquotedblleft st\textquotedblright\ and
\textquotedblleft GL\textquotedblright\ indicate the static and GL frames,
respectively. The same formula holds for the observer.

Then, 
\begin{equation}
(1+z)_{st}=\frac{\left( \omega _{st}\right) _{em}}{\left( \omega
_{st}\right) _{obs}}=(1+z)_{GL}\sqrt{\frac{1-v_{em}}{1+v_{em}}}\sqrt{\frac{%
1+v_{obs}}{1-v_{obs}}}.  \label{dop}
\end{equation}%
Taking into account (\ref{out}), we obtain

\begin{equation}
(1+z)_{st}=\frac{\sqrt{1-v_{obs}^{2}}}{\sqrt{1-v_{em}^{2}}}=\frac{\sqrt{%
f_{obs}}}{\sqrt{f_{em}}}\text{.}  \label{1+z}
\end{equation}%
Here, we reproduced the well-known result (see, e.g. Sec. 2.3.1 in \cite{fn}%
)\ which is ascribed usually to the gravitational time delay. Note that
this concept does not show up explicitly in the presented derivation, and we
have only a combination of a \textquotedblleft
cosmological\textquotedblright\ redshift with the usual SR Doppler formula.
If $r_{obs}>r_{em}$, light propagates outward, $f_{obs}>f_{em}$, $z>0$, and
we have redshift. If $r_{obs}<r_{em}$, light propagates inward, $z<0$, and
the resulting effect is blueshift.

In the static coordinate system, the interval of time $t$ between two pulses
remains the same during light propagation. As a result, the redshift is the
ratio of values of the function relating proper time with the coordinate
time $t$ at the moments of emission and detection. In this situation, it is
possible to think that the time goes slowly in regions of large $f$.
Moreover, if we imagine a space traveller who visits a vicinity of the event
horizon and then returns, he/she will be younger than his twin staying on
Earth. This is true even if the traveller has enough rocket fuel to move
towards the black hole and back very slowly, so that it is hard to attribute
the different time passed to usual Lorentz time dilation. However, it is
necessary to keep in mind that such an interpretation is connected
with a particular coordinate system rather than with spacetime itself. In
the Lemaitre coordinates, there is no gravitational time delay, and the black
hole twin paradox can be explained solely by SR Lorentz time dilation -- to
return back, the traveller must apply a huge Lorentz boost with respect to
his \textquotedblleft natural\textquotedblright\ free-falling frame.

Formula (\ref{1+z}) refers to a standard situation when both 
an observer and an emitter are at rest in a static gravitational field.
It is instructive to remind the reader that there is an analogue of this situation in
cosmology. This corresponds to the case of so-called tethered galaxies \cite%
{tethered}. In this case, the cosmological expansion is exactly compensated
due to a peculiar velocity towards an observer, so the proper distance
remains constant. However, the detected emission can be redshifted or
blueshifted depending on a cosmological equation of state. This fact has
been remarked on only recently in \cite{tethered} and shows clearly that the
cosmological redshift cannot be considered as a SR Doppler shift in an
expanding Universe. In the case of black holes for free-falling observer and
emitter, as we want to use the Lemaitre metric or its generalization we can
mimic this situation in the limit $r_{obs}\rightarrow \infty $. Then we can
approximately consider that the observer is at rest, as it is in the cosmological picture,
where an observer has zero peculiar velocity.
To have $r=const$, it
is necessary for the emitter to have a peculiar velocity which exactly
compensates the free-fall velocity. The emission will be always detected
with a redshift. Usually this situation is considered in static coordinates,
and it is interpreted in terms of gravitational time delay. However, we see
that we can use a different interpretation, saying that in Lemaitre metric
the non-zero redshift of an object at a constant proper distance has the
same nature as non-zero redshifts of tethered galaxies.
In the syncronous system one can also define the light travel distance as $D_l=c(\tau_{obs}-\tau_{em})$.
It follows from the staticity of the metric and the constancy of $r_{em}$ and $r_{obs}$ that $D_l=const$.    Therefore, an object with constant light travel distance is always
redshifted --- the property which is still valid in the FRW cosmology \cite%
{we2}.

\section{Conclusion}

In this methodological note, we proposed another interpretation of a redshift
in spherically symmetric static spacetimes, including black holes. We split
the calculation of the effect into two parts.  The first part includes
transition to the synchronous system that is formed by free-falling
particles. Here, the situation is very similar to that in cosmological FRW
spacetime.  In the second step, we return to the static system by means of
the local Lorentz boost. This enables us to consider it as a kind of the
Doppler shift and apply well-known corresponding formulas. It is worth
noting that calculations on step 1 are realized by us by two methods. We
found (i) the frequency from equations of motion, (ii) the wavelength taking
into account that different points move with different velocities (in
analogy with cosmology). Thus, we see that there is no crucial difference
between interpretations of the redshift in cosmological and black hole
spacetimes. Rather, the situation in the latter case can be thought of as a
combination of the former one and the Doppler effect.

The obtained results have a rather curious consequence. As is noticed
above, an observer crossing the horizon registers a signal from infinity
with $z=1$. This allows the observer to identify the moment of crossing the
event horizon. In doing so, an observer should use information about the
photon emitted at infinity whereas he is unable to do this using only local
measurements. (It is instructive to remind the reader that another kind of
pure local measurements can in principle locate the moment of the horizon
crossing formally, but that method is rather sophisticated and does not operate with
entities having direct meaning in physics \cite{petrov}.)

When speaking about the cosmological spacetimes throughout the paper, we
implied the FRW spacetimes as opposed to the static ones such as the
Schwarzschild metric. For completeness, we have to mention that there exist
also special cases in which cosmological spacetimes themselves admit
presentation in a static form. This includes the de Sitter and Milne
solutions when both forms (explicitly static and pure cosmological) are
possible. In \cite{melia}, six cosmological metrics with stationary frames
have been listed and the corresponding redshifts studied. It was shown that in
all these cases the cosmological redshift can be represented as a
combination of SR Doppler shift and gravitational time delay, and so it
cannot be consider as a \textquotedblleft fundamental\textquotedblright\
shift. In our paper, we did just the opposite, showing that gravitational
redshift in a stationary frame can be represented as a combination of the
Doppler shift and a "cosmological" redshift. This means that the question of
what type of redshift should be considered the more fundamental one is a
matter of convenience. On the other hand, we can argue that since a
synchronous system can be constructed for any spacetime and static systems
exist only for certain particular metrics, the \textquotedblleft
cosmological\textquotedblright\ description of a redshift presented in our
paper has more general significance.

\begin{acknowledgments}
We thank anonymous referees for their useful comments and suggestions.
This work was funded by the subsidy allocated to Kazan Federal University
for the state assignment in the sphere of scientific activities. 
The work of OBZ was  also supported by SFFR (Ukraine), project 32367.
SBP acknowledges support from the Russian Science Foundation (grant 14-12-00146).
\end{acknowledgments}

\end{document}